\documentclass[sn-mathphys-num]{sn-jnl}

\usepackage{graphicx}%
\usepackage{multirow}%
\usepackage{amsmath,amssymb,amsfonts}%
\usepackage{amsthm}%
\usepackage{mathrsfs}%
\usepackage[title]{appendix}%
\usepackage{xcolor}%
\usepackage{textcomp}%
\usepackage{manyfoot}%
\usepackage{booktabs}%
\usepackage{algorithm}%
\usepackage{algorithmicx}%
\usepackage{algpseudocode}%
\usepackage{listings}%

\begin{document}

\title[Machine Learning Approach to Predict the Curie Temperature of Fe- and Pt-Based Alloys]{Machine Learning Approach to Predict the Curie Temperature of Fe- and Pt-Based Alloys}


\author*[1]{\fnm{Svitlana} \sur{Ponomarova}}\email{sveta.ponomaryova@gmail.com}

\author[1]{\fnm{Oleksandr} \sur{Ponomarov}}\email{lxpnmrv@gmail.com}
\equalcont{These authors contributed equally to this work.}

\author[1]{\fnm{Yurii} \sur{Koval}}\email{koval@imp.kiev.ua}
\equalcont{These authors contributed equally to this work.}

\affil*[1]{\orgdiv{Phase transitions}, \orgname{Institute for Metal Physics, National Academy of Sciences of Ukraine}, \orgaddress{\street{36 Vernadsky blvd.}, \city{Kyiv}, \postcode{UA-03680}, \country{Ukraine}}}

\title{Predicting the Temperature of Magnetic Transformation in Fe-Based and Pt-Based Ternary Alloys Using Machine Learning}
\maketitle


\begin{abstract}

Various techniques can be employed to determine the temperature of magnetic transformation, whether it be the Curie or Néel temperature. The standard procedure typically involves creating alloys with defined compositions and performing measurements of $T_c$/$T_{N}$ experimentally. Alternatively, these temperatures can be predicted based on a material’s known physical and chemical properties prior to experiments.

In this study, we adopt the predictive approach and apply machine learning to estimate the Curie/Néel temperature of Fe-based and Pt-based alloys. Key design factors include atomic ordering and compositional variations introduced by adding small amounts of a third element.
We identified an optimal feature set and selected the most effective algorithm. Our findings show that the Voting Ensemble model, when combined with Monte Carlo cross-validation, achieves the highest prediction accuracy. The normalized root mean squared error serves as the primary performance metric. For implementation, we utilize the Azure Machine Learning framework for its robust computational and integration capabilities.

This approach offers an efficient and reliable strategy for designing and predicting the Curie temperature of ternary alloys.
The paper also highlights potential applications of the model and its extensions for other systems.

\end{abstract}

\section{Introduction}

The unique properties of Fe- and Pt-based alloys generate significant interest in these materials. For example, Fe-Pt alloys can experience several phase transitions, including atomic ordering, magnetic ordering, and martensitic transformation, across a wide range of concentrations and temperatures. Additionally, platinum-based alloys also undergo a magnetic transition that is associated with atomic ordering \cite{Antolini}. As noted in several studies \cite{Kakeshita}, \cite{Koval}, \cite{Ponomaryova}, different phase transitions can coexist, and their interactions lead to changes in material properties. 

Magnetic alloys, especially Fe- and Pt-based systems, are key materials for applications in data storage, spintronics, and energy-efficient technologies. 
The Curie/Néel temperature is the critical temperature at which thermal excitation is sufficient to disrupt the parallel alignment of magnetic moments, leading to the loss of spontaneous magnetization in metals and alloys. 
It can be designed by different factors, including:

\begin{itemize}
    \item \textbf{Chemical Composition}
    \begin{itemize}
        \item The type of chemical elements.
        \item Their atomic radii.
    \end{itemize}

    \item \textbf{Physical Structure}
    \begin{itemize}
        \item Type of crystal structure.
        \item Lattice parameters.
    \end{itemize}

    \item \textbf{Physical State}
    \begin{itemize}
        \item Crystalline or amorphous state.
        \item Initial thermal treatment.
        \item Atomic ordering.
    \end{itemize}

    \item \textbf{External Conditions}
    \begin{itemize}
        \item External pressure.
        \item External magnetic fields.
        \item Irradiation.
    \end{itemize}
\end{itemize}

Different methods can help to determine the temperature of magnetic transformation, such as measuring electrical resistance under high pressure \cite{Oomi}, examining changes in magnetization with temperature \cite{Nataf}, and assessing AC susceptibility under pressure \cite{Matsushita}, among other techniques.
In addition to experimental methods, various computational studies focus on calculating the Curie and Néel temperatures. For instance, these temperatures can be computed using first-principles density functional theory (DFT) \cite{Arale}. Additionally, a semi-phenomenological approach has been proposed for Fe-X alloys in \cite{Ponomarova}, which combines experimental data with analytical expressions.
However, experimental measurements and classical computations across different chemical concentrations and crystal structures might be time-consuming and resource-intensive.

Machine learning (ML) plays an important role in materials science and has already been applied to the study of magnetic properties \cite{Evan}, \cite{Jung}, \cite{Nelson}, \cite{Singh}, \cite{Lin}.

The predictions of magnetic properties facilitated by ML have significant implications in electric transport \cite{Kipp}, next-generation networks \cite{Pollok}, the Internet of Things (IoT) \cite{Milyutin}, and medicine \cite{Takiguchi}.

\subsection{Motivation}
Recent advances in machine learning have demonstrated great potential for predicting magnetic properties of metals and alloys. Many of existing studies, however, have concentrated on binary alloys and frequently neglect the influence of atomic ordering, despite its significant role in magnetic behavior. This creates a critical gap in predictive modeling relevant for ternary systems.

To address this challenge, we propose a machine learning pipeline tailored for three-component alloys, using Fe- and Pt-based ternary alloys as representative examples. By incorporating both chemical composition and atomic ordering as design factors, our approach achieves accurate predictions of Curie temperature. The pipeline offers a ready-to-use feature set and algorithmic framework that can be applied to other classes of magnetic materials, accelerating data-driven discovery in magnetism and alloy design.

Our investigation focuses on Fe-based and Pt-based ternary alloys. This work is motivated by previous studies that have applied machine learning techniques to predict the magnetic properties of metals and alloys \cite{Arale}, \cite{Jung}, \cite{Belot}, \cite{Ajaib}. We developed a comprehensive pipeline starting with a dataset derived from publicly available experimental data. Multiple machine learning algorithms were evaluated, and their parameters systematically optimized.

\subsection{Contribution}

This paper presents a machine learning framework for estimating the Curie temperature of ternary alloys, thereby advancing the application of data-driven approaches in materials science. Building upon prior studies on binary systems such as Fe–Pt, Fe–Pd, Co–Pt, and Fe–Ni \cite{Ponomarova_ML}, this work extends the methodology to more complex ternary alloys. A central contribution of this study is our approach to predicting the temperature of magnetic transformation in ternary alloys, which explicitly considers both composition and atomic ordering as key design parameters. In particular, it systematically explores how variations in chemical composition and atomic ordering can be incorporated as key design parameters for predicting the Curie temperature.

\subsection{Paper Structure}
The paper is structured as follows. Section 2 introduces some ML terminology used, detailing the selection of feature datasets, the validation metrics, the algorithms employed, and the tuning parameters utilized to enhance prediction accuracy. Section 3 presents the results of predicting the Curie or Néel temperature for the Fe-Pt-Pd ternary system, exploring various concentrations of palladium as it replaces either iron (Fe) or platinum (Pt) atoms. This section also analyzes the dependence of the Curie/Néel temperature on alloy composition and the degree of atomic order. Then we discuss the long-range order parameter and chemical composition from a thermodynamic perspective, and outline potential model applications.

\subsection{Related works}
The field of predicting magnetic transformation temperatures in various materials is extensive and encompasses a range of techniques. Table \ref{table:ml_Curie} presents recent research focused on predicting Curie temperatures using machine learning methods, though it is not limited to the cited works.

\begin{table}[ht]%
\caption{\label{table:feature} Summary of related papers focused on predicting Curie temperatures using machine learning}
\begin{tabular*}{\textwidth}{@{}l*{15}{@{\extracolsep{0pt plus
12pt}}l}}
\hline
Short work summary    & Citation \\

\hline
\verb\\  Predicting the Curie temperature of magnetic materials with automated calculations & \cite{Arale}  \\
\verb\\ Machine-Learning prediction of Curie temperature from chemical compositions  & \cite{Jung}  \\
\verb\\ Discovery of functional magnetic materials with ML  & \cite{Singh}  \\
\verb\\ ML-aided framework for rapid discovery of two-dimensional ferromagnets  & \cite{Lu111}  \\
\verb\\ Machine learning-based Curie temperature prediction for magnetic 14:2:1 phases  & \cite{Kumar}  \\
\verb\\ Explainable AI-supported interpretation for Curie temperature in Heusler alloys  & \cite{Hilgers}  \\
\verb\\ Machine learning predictions of high Curie-temperature materials  & \cite{Belot}  \\
\verb\\ An accelerating ML approach to design materials by modeling magnetic ground state  & \cite{Long}  \\
\end{tabular*}
\label{table:ml_Curie}
\end{table}

\section{Computational technique}

\begin{figure}[H]%
\centering
\includegraphics[width=\textwidth]{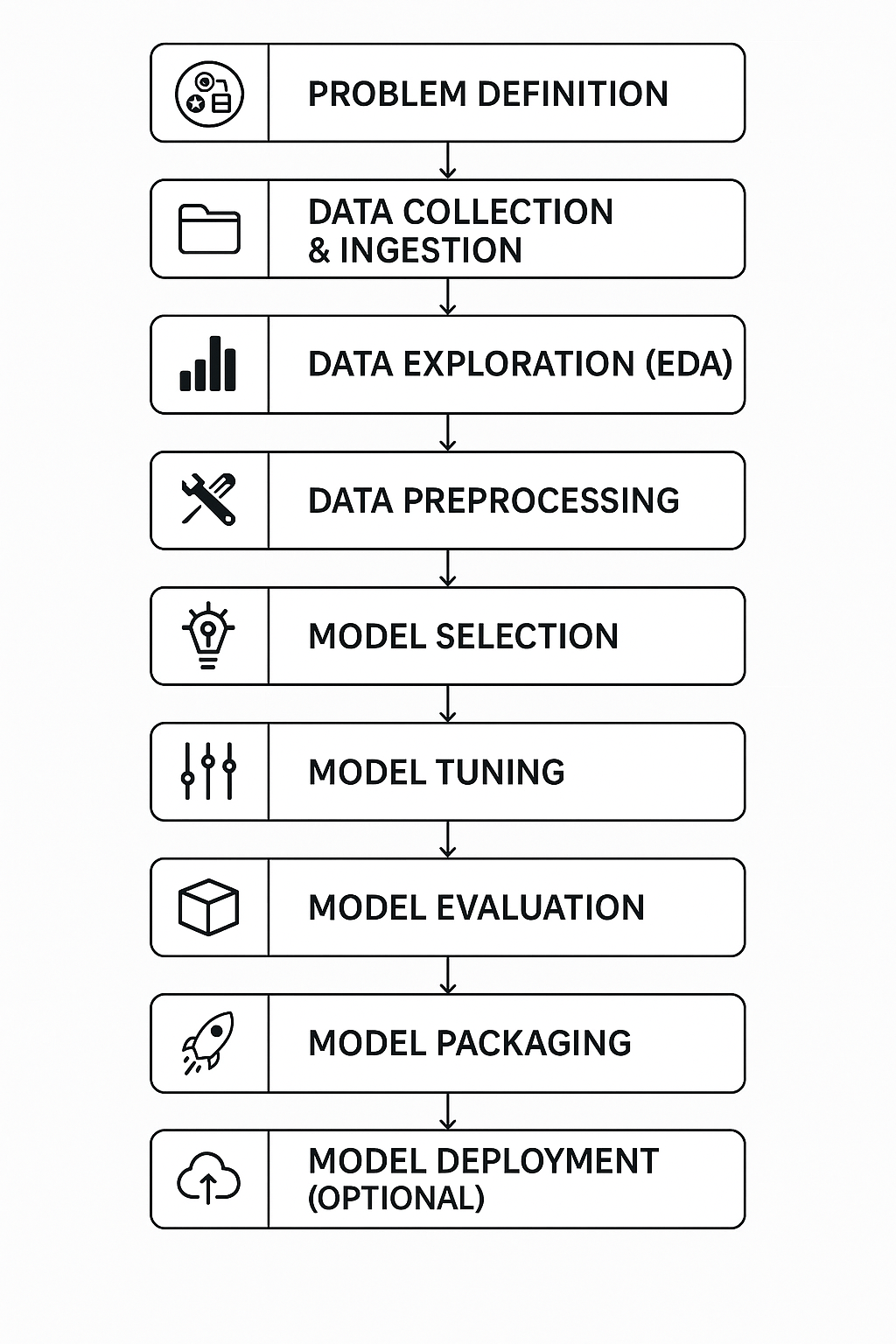}
\caption{Operational Workflow Overview for ML Models}\label{fig:model}
\end{figure}

As a technical tool, we use Azure Machine Learning Studio \footnote{https://learn.microsoft.com/en-us/azure/machine-learning/tutorial-azure-ml-in-a-day?view=azureml-api-2}, which leverages cloud computing capabilities. 
Given that ML models are frequently viewed as "black boxes," we will briefly outline the typical stages involved (Fig. \ref{fig:model}).

\textit{Problem definition.} 
The core objective of this research is to develop a machine learning model capable of predicting the Curie/Néel temperature in Fe-based and Pt-based three-component alloys, with potential applicability to other ternary alloys.

\textit{Data Collection and Ingestion.}
Feature selection is one of the most important and challenging tasks in machine learning. We examined the characteristics of Fe-Pt and subsequently engineered a high-dimensional feature vector: $f(c_{Fe}, c_{Pt},c_{Pd},r_{Fe},r_{Pt}, r_{Pd},s_{Fe},s_{Pt},s_{Pd},Z_{Fe},Z_{Pt},Z_{Pd},\eta)$ with the features summarized in Table \ref{table:features}. The manual selection criteria based on the essential material properties is applied to define the feature vector.
It includes the concentrations of Fe ($c_{Fe}$), Pt ($c_{Pt}$), and Pd ($c_{Pd}$) along with their spin numbers: $s_{Fe}$, $s_{Pt}$, $s_{Pd}$,
the atomic radii ($r_{Fe}$, $r_{Pt}$, $r_{Pd}$), and atomic numbers $Z_{Fe}$, $Z_{Pt}$, $Z_{Pd}$ which will be important while extending this model for other multicomponent alloys. The Curie/Néel temperature ($T_c$/$T_N$) is also included into feature vector.

\begin{table}[ht]%
\caption{\label{table:features} A List of features taken for Curie temperature prediction in Fe-Pt-Pd}
\begin{tabular*}{\textwidth}{@{}l*{15}{@{\extracolsep{0pt plus
12pt}}l}}
\hline
Feature      & Feature description \\
\hline
\verb\\$c_{Fe}$ & concentration of Fe  \\
\verb\\$c_{Pt}$ & concentration of Pt   \\
\verb\\$c_{Pd}$ & concentration of Pd   \\
\verb\\$r_{Fe}$ & atomic radius of Fe  \\
\verb\\$r_{Pt}$ & atomic radius of Pt  \\
\verb\\$r_{Pd}$ & atomic radius of Pd  \\
\verb\\$s_{Fe}$ & spin number of Fe  \\
\verb\\$s_{Pt}$ & spin number of Pt  \\
\verb\\$s_{Pd}$ & spin number of Pd  \\
\verb\\$Z_{Fe}$ & atomic number of Fe  \\
\verb\\$Z_{Pt}$ & atomic number of Pt  \\
\verb\\$Z_{Pd}$ & atomic number of Pd  \\
\verb\\$\eta$ & long-range order parameter \\
\end{tabular*}
\end{table}

Palladium (Pd) is known to be a non-magnetic element \cite{Polesya}. However, like platinum (Pt), it can acquire a magnetic moment when influenced by magnetic iron (Fe) atoms.
Direct experiments have demonstrated the induced magnetism of platinum (Pt) and other 5d elements when interacting with 3d ferromagnets \cite{Poulopoulos}. This fact highlights the importance of including the spin moments of Pd and Pt into the feature vector.

The input data is collected from published experimental papers \cite{Oomi}, \cite{Kakeshita2}, \cite{Koval2}, \cite{Yamamoto}, \cite{Sumiyama}, \cite{Mizoguchi}, \cite{Barmak} and publicly available
databases and resources, such as the Materials Project \footnote{https://next-gen.materialsproject.org} and AFLOW \footnote{https://aflowlib.org}.
A sample of the dataset is presented in Table \ref{table:dataset}. 

\begin{table}[ht]%
\caption{\label{table:dataset} Dataset sample}
\begin{tabular*}{\textwidth}{@{}l*{15}{@{\extracolsep{0pt plus
12pt}}l}}
\hline
$c_{Fe}$ & $c_{Pt}$ & $c_{Pd}$ & $r_{Fe}$ & $r_{Pt}$ & $r_{Pd}$ & $s_{Fe}$ & $s_{Pt}$ & $s_{Pd}$ & $Z_{Fe}$ & $Z_{Pt}$ & $Z_{Pd}$ & $\eta$ & $T_C$/$T_N$ \\
\hline
\hline
\verb"76"& 24 & 0 & 126 & 139 & 137 & 2.73 & 0.38 & 0.289 & 26 & 78 & 46 & 0.5 & 288\\
\verb"76"& 24 & 0 & 126 & 139 & 137 & 2.73 & 0.38 & 0.289 & 26 & 78 & 46 & 0.8 & 333\\
\verb"75"& 25 & 0 & 126 & 139 & 137 & 2.73 & 0.38 & 0.289 & 26 & 78 & 46 & 0 & 270\\
\verb"75"& 25 & 0 &126 & 139 & 137 & 2.73 & 0.38 & 0.289 & 26 & 78 & 46 & 0.57 & 353\\
\verb"75"& 25 & 0 &126 & 139 & 137 & 2.73 & 0.38 & 0.289 & 26 & 78 & 46 & 0.88 & 441\\
\end{tabular*}
\end{table}

\textit{Data Preprocessing} includes analysis of the input dataset with different algorithms:
\begin{itemize}
    \item Standard Scaler Wrapper \footnote{https://scikit-learn.org/stable/modules/generated/sklearn.preprocessing.StandardScaler.html}
    \item Robust Scaler \footnote{https://scikit-learn.org/stable/modules/generated/sklearn.preprocessing.RobustScaler.html}
    \item Min Max Scaler \footnote{https://scikit-learn.org/stable/modules/generated/sklearn.preprocessing.MinMaxScaler.html}
\end{itemize}
We handled the missing values using the median imputation method, which is less sensitive to extreme values than the mean imputation method \cite{Alam}.
Data guardrails help detect potential issues in the input dataset and correct inaccuracies by performing a series of validation checks to ensure that only high-quality data are used for model training.

\textit{Model Selection.}
We explore supervised machine learning techniques. A series of machine learning experiments has been designed and performed in Azure ML Studio using the following algorithms:

\begin{itemize}
    \item Voting Ensemble \cite{Abro}
    \item Extreme Random Trees \cite{Geurts}
    \item Random Forest \footnote{https://scikit-learn.org/stable/modules/generated/sklearn.ensemble.RandomForestClassifier.html}
    \item Gradient Boosting \footnote{https://scikit-learn.org/stable/modules/generated/sklearn.ensemble.GradientBoostingClassifier.html}
\end{itemize}

For all of them, the Azure ML metrics \footnote{https://learn.microsoft.com/en-us/azure/machine-learning/how-to-understand-automated-ml?view=azureml-api-2} are analyzed. The best predicting results in Fe-Pt-Pd alloys are shown by the Voting Ensemble algorithm,  where the final prediction is made by combining the individual predictions using a voting scheme.

\textit{Model Tuning.}
For a regression model, the Azure Machine Learning framework requires a dataset with a minimum of 50 rows to initiate training.  Since the dataset is derived from experimental measurements, its size remains relatively limited (our dataset contains 100 rows) compared to typical machine learning requirements. In the context of machine learning, this illustrates a common challenge often specific to materials science — the problem of small data, discussed in \cite{Xu}. 
The "small data" issue can be resolved by adjusting tuning parameters and using the appropriate machine learning model.
As tuning parameters, we employ L1-norm and L2-norm regularization, following the approach proposed in \cite{Luo}, to improve model generalization and reduce the risk of overfitting.

The trained model was validated using Monte Carlo cross-validation, a method thoroughly explained in the work by Xu et al \cite{Xu2}. It is well-suited for small datasets and helps to address variance and bias concerns. The Monte Carlo cross-validation uses repeated random sampling to average out the variability and results in more reliable performance metrics. 

\textit{Model Evaluation.}

Table \ref{table:metrics} presents the performance metrics of the best-performing algorithm.

\begin{table}[ht]%
\caption{\label{table:feature} A List of metrics for Voting Ensemble algorithm}
\begin{tabular*}{\textwidth}{@{}l*{15}{@{\extracolsep{0pt plus
12pt}}l}}
\hline
Metric      & Value \\
\hline
\verb\\Explained variance  & 0.927  \\
\verb\\Mean absolute error & 29.656 \\
\verb\\Mean absolute percentage error  & 7.672  \\
\verb\\Median absolute error   & 11.680  \\
\verb\\Normalized mean absolute error   & 0.028  \\
\verb\\Normalized median absolute error  & 0.011  \\
\verb\\Normalized root mean squared error & 0.051  \\
\verb\\Normalized root mean squared log error  & 0.046  \\
\verb\\R2 score    & 0.903  \\
\verb\\Root mean squared error     & 52.399  \\
\verb\\Root mean squared log error & 0.319  \\
\verb\\Spearman correlation        & 0.960  \\
\end{tabular*}
\label{table:metrics}
\end{table}

We use the Normalized Root Mean Squared Error as the primary metric for the model accuracy, comparing predicted values with observed values and normalizing the result by the range of the observed data. We achieved an NRMSE of 0.051, which is relatively low, indicating a good correlation between the model's predictions and the observed experimental data.

\textit{Model Packaging and Deployment.}
A prepared and validated model can be packaged and deployed to estimate temperatures of magnetic transformation in ternary alloys, supporting fast, data-driven materials design.

\section{Results and Discussion}
\subsection{Object of research}

In this study, the Fe–Pt alloy system is selected as the base composition, with palladium (Pd) introduced as a third element (denoted as X) in concentrations up to 5 atomic percent (at.\%). The investigation focuses on the substitutional effects of Pd when replacing either iron (Fe) or platinum (Pt) atoms within the Fe–Pt lattice. Accordingly, the following compositional variants are considered:
\newline
* $Fe_{75}Pt_{25-X}Pd_X$
\newline
* $Fe_{75-X}Pt_{25}Pd_X$
\newline
* $Fe_{25}Pt_{75-X}Pd_X$
\newline
* $Fe_{25-X}Pt_{75}Pd_X$

The study specifically targeted alloys with the $L1_2$-type stoichiometry, as this phase enables systematic control of the degree of atomic ordering— an essential parameter in tailoring the magnetic characteristics of the system.

\subsection{Prediction of Curie temperature in Fe-Pt-Pd}
Experimental research \cite{Oomi} shows a correlation between magnetic transition temperature and atomic ordering in Fe-Pt compounds. Therefore, we investigated both disordered and ordered states for ternary Fe-Pt-Pd alloys.

\begin{figure}[ht]%
    \centering
    \includegraphics[width=\linewidth]{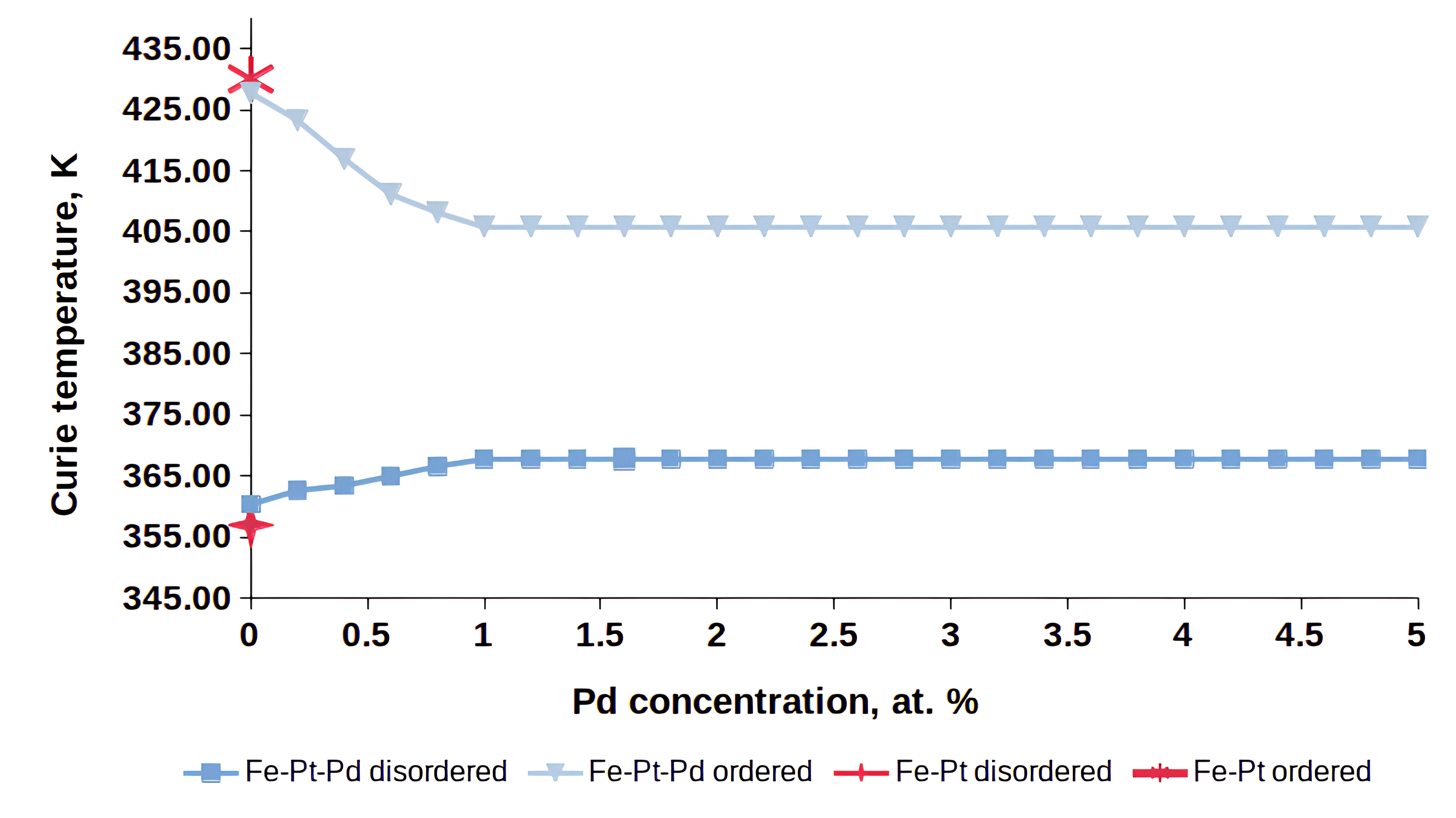}
    \caption{Curie temperature for $Fe_{75}Pt_{25-X}Pd_X$ where Pd replaces Pt atoms}
    \label{fig:1}
\end{figure}

\begin{figure}[H]%
    \centering
    \includegraphics[width=\linewidth]{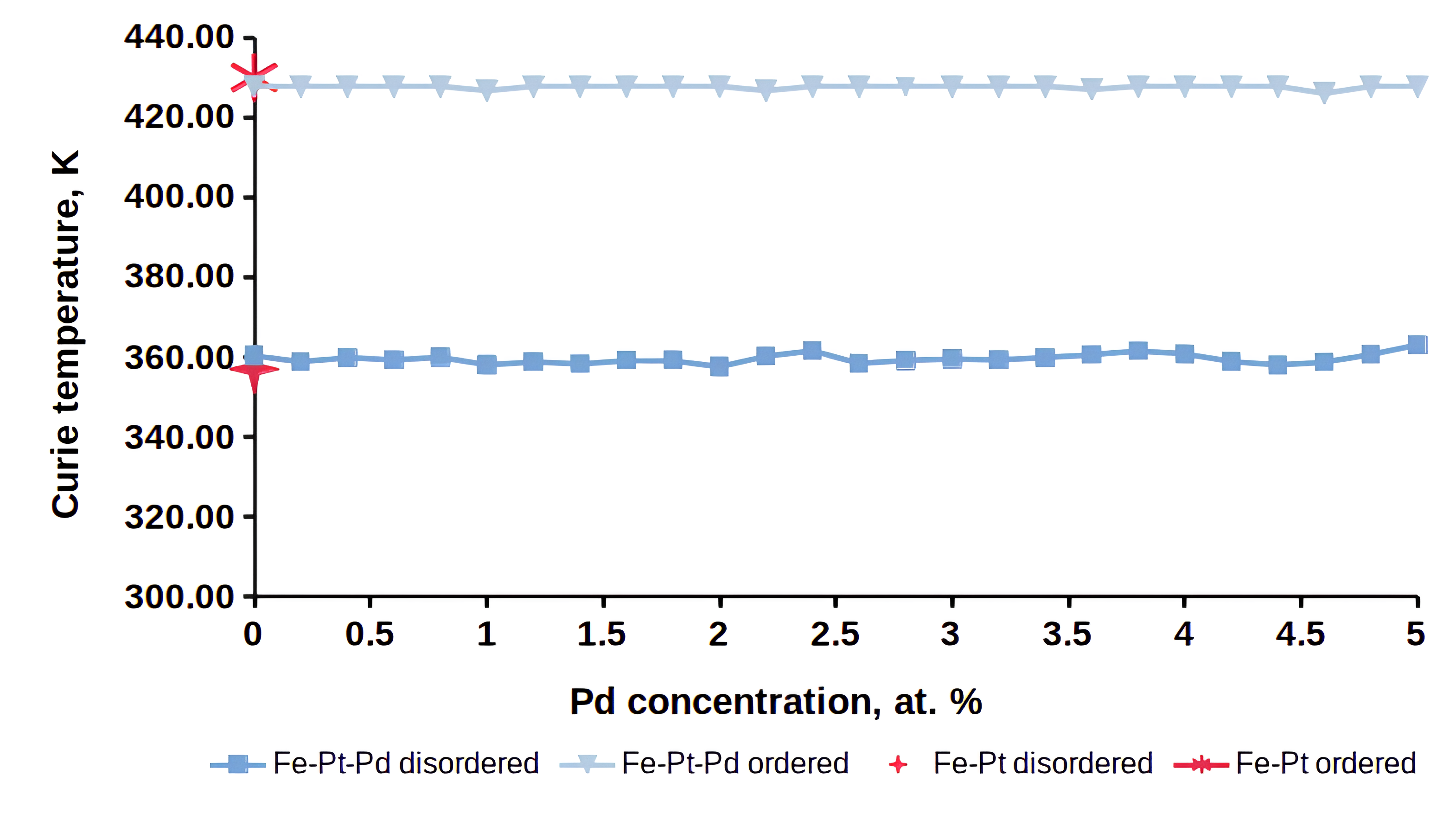}
    \caption{Curie temperature for $Fe_{75-X}Pt_{25}Pd_X$ where Pd replaces Fe atoms}
    \label{fig:2}
\end{figure}

Figures \ref{fig:1} and \ref{fig:2} show the Curie temperature values predicted by the ML model for Fe-rich alloys.
Introducing a small amount of palladium (Pd) into the $Fe_{75}Pt_{25-X}Pd_X$ alloys affects the Curie temperature in different ways, depending on the alloy's initial state. For the ordered state, adding Pd decreases the Curie temperature, while for the disordered state, it increases. Up to 1 atomic percent of Pd the Curie temperature rises slightly, by about 10 K for the disordered alloy, and it is decreased by 25 K for the ordered alloy. Increasing the palladium content beyond 1 atomic percent does not further change the Curie temperature; it levels off at an asymptotic value in both the ordered and disordered states.
In the case of substituting Fe atoms with Pd in the $Fe_{75-X}Pt_{25}Pd_X$ alloys, no change in the Curie temperature is observed for either the ordered or disordered states (see Fig. \ref{fig:2}).

Experimentally measured Curie temperature for binary $Fe_3Pt$ alloy in disordered state is 353 K \cite{Yamamoto}, while Curie temperature in ordered state is 430 K \cite{Sumiyama}.
As seen, the experimental values for binary Fe–Pt alloys (marked in red in Fig. \ref{fig:1}, Fig. \ref{fig:2}) closely match the predictions, confirming the validity of the trained model.

\begin{figure}[H]%
    \centering
    \includegraphics[width=0.37\linewidth]{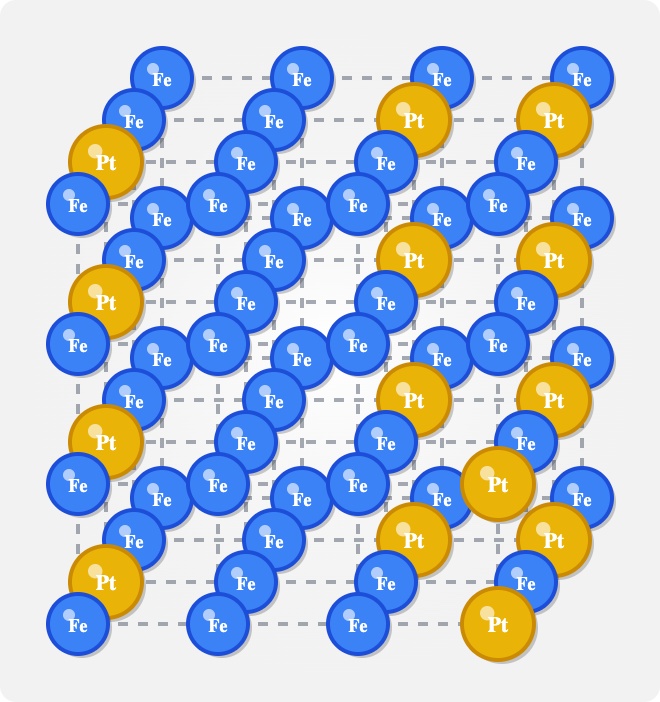}
    \caption{Disordered state for Fe-Pt alloys}
    \label{fig:dis}
\end{figure}

\begin{figure}[H]%
    \centering
    \includegraphics[width=0.37\linewidth]{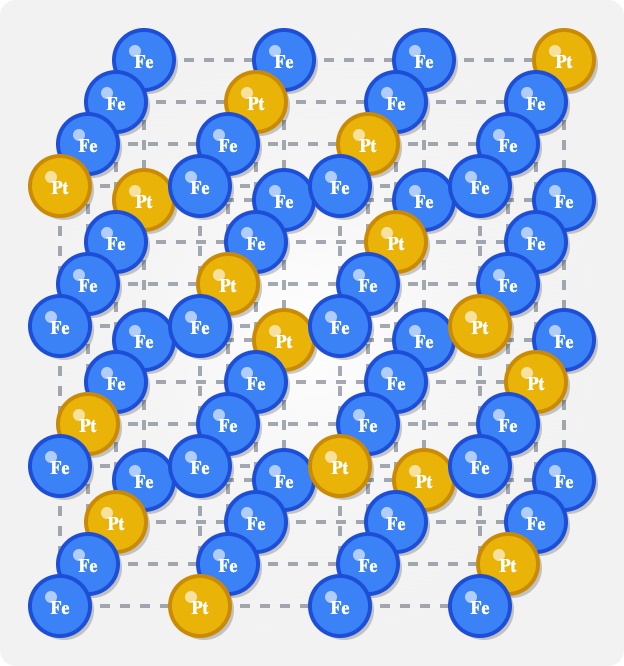}
    \centering
    \caption{Atomic ordering in Fe-Pt alloys}
    \label{fig:ord}
\end{figure}

\begin{figure}[H]%
    \centering
    \includegraphics[width=0.37\linewidth]{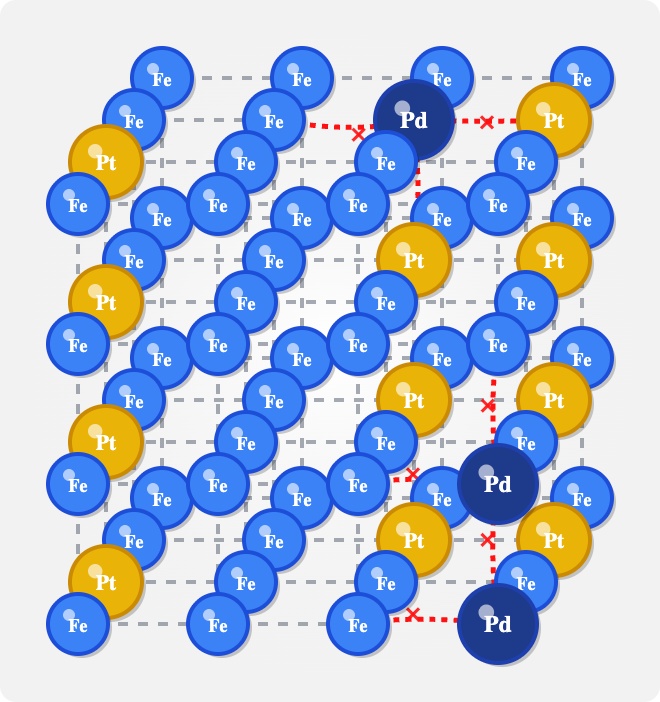}
    \caption{Positional disorder in Fe-Pt-Pd when alloying Fe-Pt by Pd}
    \label{fig:positional}
\end{figure}

The general observation for Fe-rich alloys is that the Curie temperature in the ordered state with a predominant near-coordination environment (Fig. \ref{fig:ord}) is significantly higher than in the disordered state with the random location of atoms (Fig. \ref{fig:dis}) \cite{Yamamoto}, \cite{Sumiyama}.

According to Madelung's terminology \cite{Madelung}, alloying introduces positional disorder in solid-state materials, increasing internal tension, as shown with strain bonds (marked by red color) for Pd atoms in Fig. \ref{fig:positional}.
Therefore, in Fe-Pt-Pd alloys, we see two competing factors at play: atomic ordering and positional disorder. 
Atomic ordering is characterized by the $L1_2$ structure, which results from annealing and is described by the long-range order parameter $\eta$. On the other hand, positional disorder arises from the incorporation of Pd interstitials in the alloy and is described by $c_{Pd}$. The thermodynamic state of the alloy can be described by $f(c_{Pt}, c_{Pd}, \eta)$.

The binary Fe-Pt alloys show a ferromagnetic type of phase transition with a Curie temperature higher in the atomically ordered state in comparison to the disordered state \cite{Yamamoto}, \cite{Sumiyama}.
A small amount of palladium, which replaces Pt, retains the same tendency and keeps the Curie temperature in disordered three-component alloys lower than in ordered ones (Fig. \ref{fig:1}), indirectly pointing to a ferromagnetic type of phase transition in Fe-rich Fe-Pt-Pd alloys.

\begin{figure}[H]%
    \includegraphics[width=\linewidth]{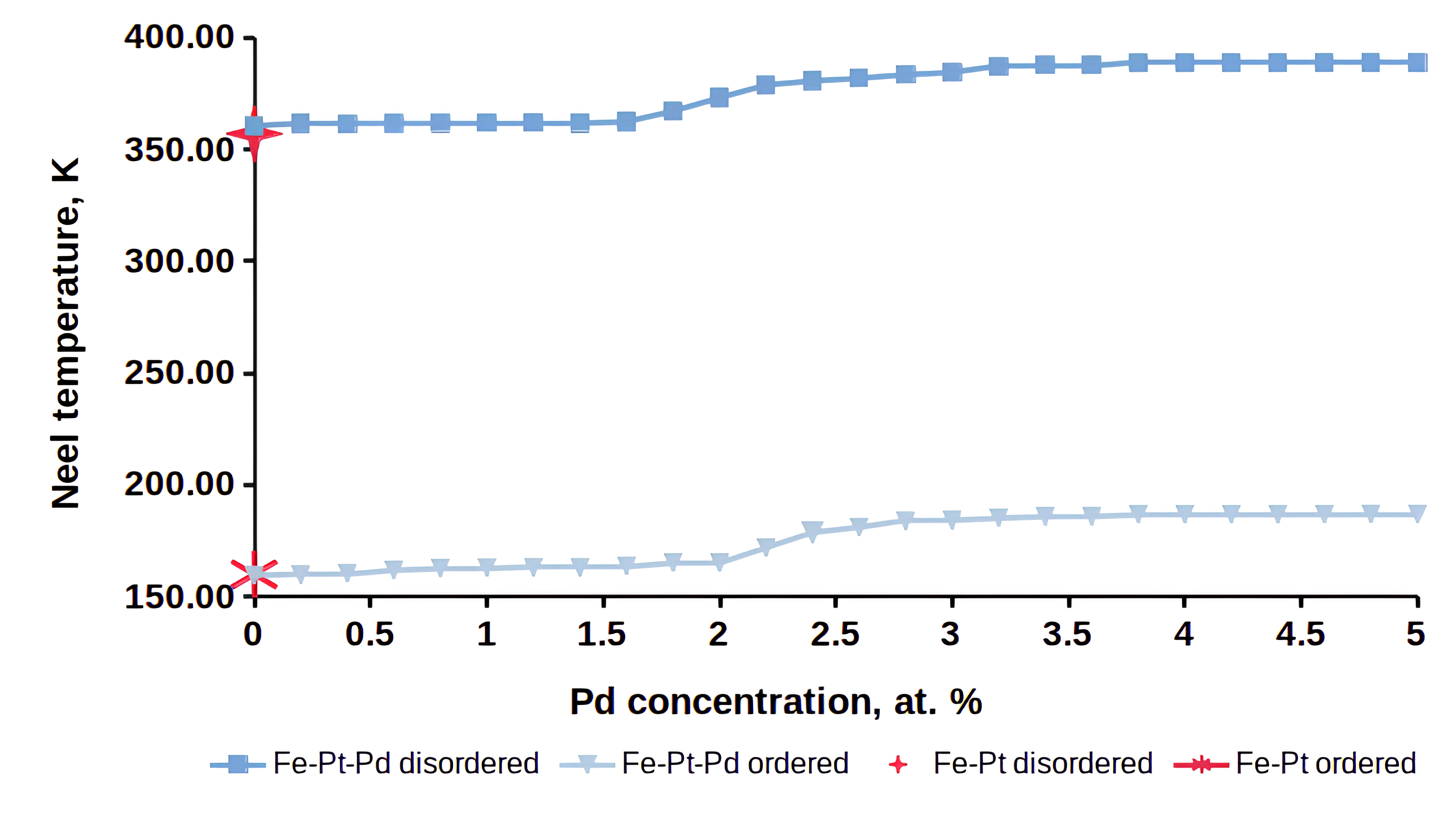}
    \caption{Néel temperature for $Fe_{25}Pt_{75-X}Pd_X$ where Pd replaces Pt atoms}
    \label{fig:3}
\end{figure}

\begin{figure}[ht]%
    \includegraphics[width=\linewidth]{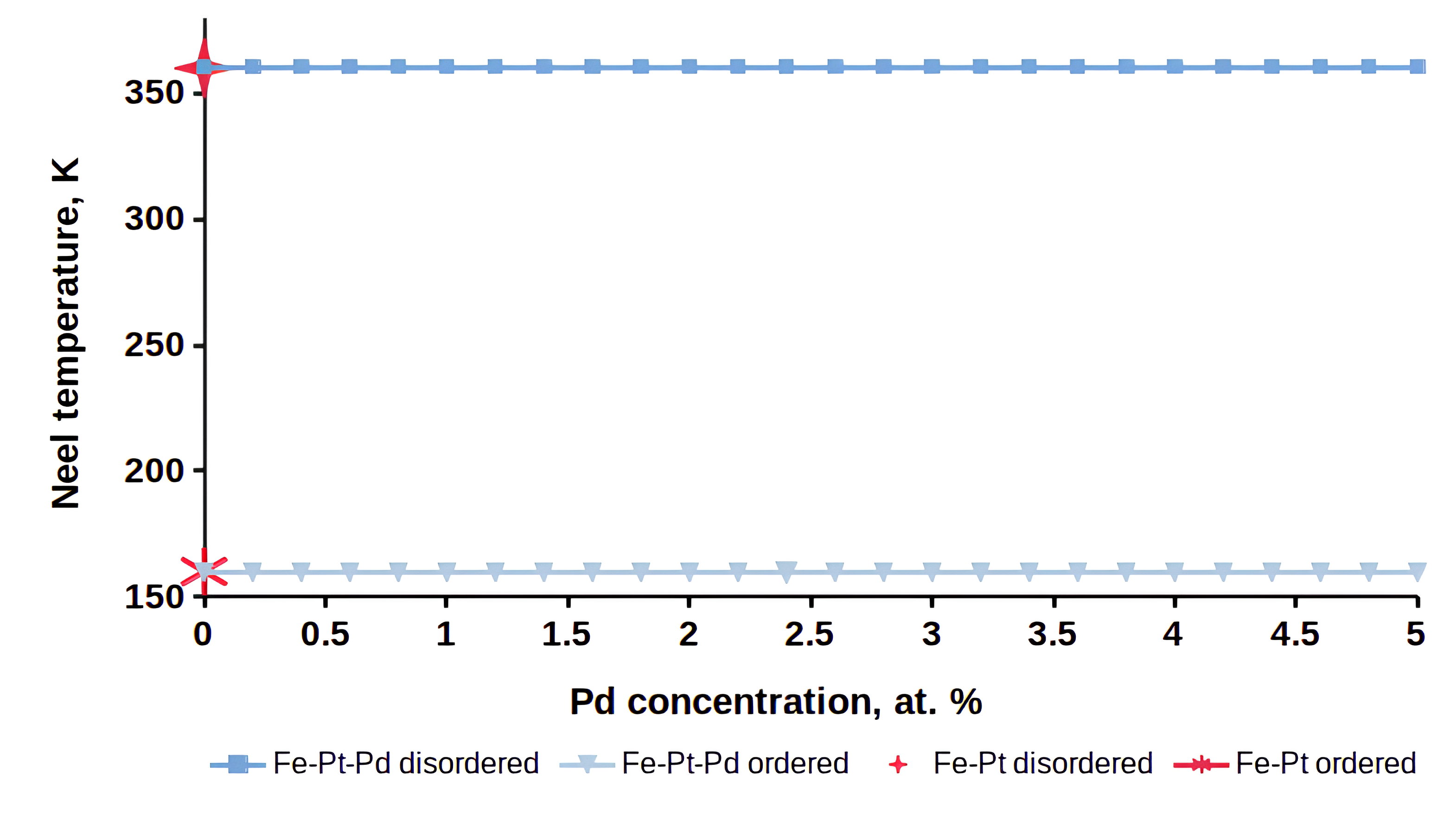}
    \caption{Néel temperature for $Fe_{25-X}Pt_{75}Pd_X$ where Pd replaces Fe atoms}
    \label{fig:4}
\end{figure}

Let us examine the behavior of the magnetic transition temperature in the case of Pt-rich alloys.
We observe a slight dependence of critical temperature on concentration when platinum atoms are replaced with palladium (Fig. \ref{fig:3}). Interestingly, after approximately 2 atomic percent (at.\%) of platinum is replaced with palladium, the value reaches an asymptotic limit. However, there is no change when Fe atoms are replaced with Pd, as illustrated in Fig. \ref{fig:4}.
Similar behavior is also observed for Fe-rich alloys in Fig. \ref{fig:1}, Fig. \ref{fig:2}.

Binary $FePt_3$ alloys display a distinctive atomic order, and their magnetic structure changes with temperature. A key feature of ordered, platinum-rich Fe–Pt alloys is their antiferromagnetic behavior, which is closely related to the degree of atomic order \cite{Bacon}. In particular, alloys with a high degree of ordering tend to display an antiferromagnetic structure.

Experimental investigations on Pt-rich binary alloys \cite{Tobita}, \cite{Mani} have shown that the formation of $L1_2$ type ordered structure in $FePt_3$ leads to a magnetic transition from ferromagnetic to antiferromagnetic behavior.
Moreover, the Néel temperature of $FePt_3$ alloys is significantly lower \cite{Mani}, \cite{HeitschDoh} than the Curie temperature observed in their disordered state \cite{Yin}.

As seen in Figs. \ref{fig:3} and \ref{fig:4}, such a shift also appears to be present in the platinum-rich Fe-Pt-Pd alloys, which are the subject of the current investigation. The results on the magnetic transformation temperature may indirectly confirm the existence of an antiferromagnetic state in atomically ordered Fe-Pt-Pd alloys.

\subsection{Thermodynamic analysis}
In this study, we treat alloy composition and the degree of atomic order as the principal design parameters to define the magnetic transition temperature (Curie or Néel temperature). The thermodynamic state of the binary magnetic system is described by a configuration‑dependent free energy per lattice site; the explicit form of this free energy is explained in \cite{Ponomarova}.

For the three‑component system, the corresponding relation becomes more complex. It has to be generalized to account for the additional compositional degrees of freedom and the larger space of atomic configurations.
Consequently, when extending from a binary to a three‑component alloy, the free‑energy expression is changed to include cross‑interaction terms and the additional configurational variables.
\begin{gather}
    f(c_{Fe},c_{Pt}, c_{Pd}, \eta) =  
    \frac{1}{2}  \bigg [ \bigg (
       c^2_{Pd} \tilde{w}_{prm}(\textbf{0})
       + c_{Fe}^2\tilde{J}_{FeFe}(\textbf{0})s^2_{Fe}\sigma^2_{Fe}
       \nonumber \\
       + c^2_{Pt}\tilde{J}_{PtPt}(\textbf{0})s^2_{Pt}\sigma^2_{Pt}
       + c^2_{Pd}\tilde{J}_{PdPd}(\textbf{0})s^2_{Pd}\sigma^2_{Pd}
       + 2 c_{Fe} c_{Pt} \tilde{J}_{FePt}(\textbf{0}) s_{Fe} \sigma_{Fe} s_{Pt} \sigma_{Pt} 
       \nonumber \\
       + 2 c_{Fe} c_{Pd}\tilde{J}_{FePd}(\textbf{0}) s_{Fe} \sigma_{Fe} s_{Pd} \sigma_{Pd} 
       + 2 c_{Pt} c_{Pd}\tilde{J}_{PtPd}(\textbf{0}) s_{Pt} \sigma_{Pt} s_{Pd} \sigma_{Pd}\bigg )
       \nonumber \\
      + \frac{3}{16}\eta^2 \bigg (
            \tilde{w}_{prm}(\textbf{k}_x) +
            \tilde{J}_{PtPt}(\textbf{k}_x)s^2_{Pt}\sigma^2_{Pt} 
            \nonumber
            + \tilde{J}_{FeFe}(\textbf{k}_x)
            s^2_{Fe}\sigma^2_{Fe}
            - 2\tilde{J}_{FePt}(\textbf{k}_x)s_{Fe}s_{Pt}\sigma_{Fe}\sigma_{Pt}  
            \\
            + \tilde{J}_{PdPd}(\textbf{k}_x)s^2_{Pd}\sigma^2_{Pd}
            - 2\tilde{J}_{FePd}(\textbf{k}_x)s_{Fe}s_{Pd}\sigma_{Fe}\sigma_{Pd}
            - 2\tilde{J}_{PtPd}(\textbf{k}_x)s_{Pt}s_{Pd}\sigma_{Pt}\sigma_{Pd} \bigg ) \bigg ]
            \nonumber \\
           + \frac{k_{B}T}{4}
            \bigg [
                3 \bigg (c_{Pt} - \frac{\eta}{4} \bigg )ln \bigg (c_{Pt} - \frac{\eta}{4} \bigg )
                +
                3 \bigg (c_{Fe} + \frac{\eta}{4} \bigg)ln\bigg (c_{Fe} + \frac{\eta}{4} \bigg) \\
                +
                \bigg (c_{Fe} - \frac{3\eta}{4} \bigg )ln \bigg ( c_{Fe} - \frac{3 \eta}{4} \bigg) \nonumber
                +
                \bigg (c_{Pt} + \frac{3\eta}{4} \bigg)
                ln \bigg (c_{Pt} + \frac{3\eta}{4} \bigg ) \bigg] 
                \nonumber \\
                - k_{B}Tc_{Fe}
                \bigg [lnsh \bigg(
                        \bigg (
                            1 + 
                            \frac{1}{2s_{Fe}}
                        \bigg )
                        y_{Pt}(\sigma_{Fe})
                    \bigg ) \nonumber
                    -
                    lnsh \bigg (\frac{1}{2s_{Fe}}y_{Fe}(\sigma_{Fe}) \bigg)
                    -
                    \sigma_{Fe}y_{Fe}(\sigma_{Fe} ) \bigg] 
                    \nonumber \\
                    - k_{B}Tc_{Pt}
                \bigg [lnsh \bigg(
                        \bigg (
                            1 + 
                            \frac{1}{2s_{Pt}}
                        \bigg )
                        y_{Pt}(\sigma_{Pt})
                    \bigg ) \nonumber
                    -
                    lnsh \bigg (\frac{1}{2s_{Pt}}y_{Pt}(\sigma_{Pt}) \bigg)
                    -
                    \sigma_{Pt}y_{Pt}(\sigma_{Pt} ) \bigg]
                    \nonumber \\
                    - k_{B}Tc_{Pd}
                \bigg [lnsh \bigg(
                        \bigg (
                            1 + 
                            \frac{1}{2s_{Pd}}
                        \bigg )
                        y_{Pd}(\sigma_{Pd})
                    \bigg ) \nonumber
                    -
                    lnsh \bigg (\frac{1}{2s_{Pd}}y_{Pd}(\sigma_{Pd}) \bigg)
                    -
                    \sigma_{Pd}y_{Pd}(\sigma_{Pd} ) \bigg]
        \label{free_en_3}
\end{gather}
\linebreak
Here, $\tilde{J}_{FeFe}$, $\tilde{J}_{FePt}$, $\tilde{J}_{PtPt}$, $\tilde{J}_{FePd}$, $\tilde{J}_{PtPd}$, $\tilde{J}_{PdPd}$ are the Fourier components of the exchange coupling constants corresponding to the atomic pairs Fe-Fe, Fe-Pt, Fe-Pd, Pt-Pt, Pt-Pd, and Pd-Pd, respectively.

Configurational free energy includes relative spontaneous magnetizations ($\sigma_{Fe}$, $\sigma_{Pt}$, $\sigma_{Pd}$);
concentrations ($c_{Fe}$, $c_{Pt}$, $c_{Pd}$) of Fe, Pt, and Pd; and their spin numbers ($s_{Fe}$, $s_{Pt}$, $s_{Pd}$).

For the three-component system considered, the total concentration of the alloy is given by:  $c_{Fe} + c_{Pt} + c_{Pd} = 1$, where $c_{Pd}$ $<<$ $c_{Fe} + c_{Pt}$. So, relation $f(c_{Fe},c_{Pt}, c_{Pd}, \eta)$ can be transformed to $f(c_{Pt}, c_{Pd}, \eta)$, having only independent degrees of freedom.

The Fourier component of  ‘paramagnetic' ‘mixing’ energy for components are initialized by $\tilde{w}_{(Fe-Pt-Pd)}(\textbf{k}_x)$, $\tilde{w}_{(Fe-Pt)}(\textbf{k}_x)$, $\tilde{w}_{(Fe-Pt-Pd)}(\textbf{0})$, and $\tilde{w}_{(Fe-Pt)}(\textbf{0})$. 

The function $y_{\alpha}(\sigma_{\alpha})=s_{\alpha} H_{eff}^{\alpha}/{k_B T} << 1$, where $\alpha \in {Fe,Pt,Pd}$, is defined by spin number $s_{\alpha}$ the relative spontaneous magnetization for each alloy component with the effective self-consistent ('molecular') field $H_{eff}^{\alpha}$.

It is essential to note that the sign of the exchange coupling constants depends on the type of magnetic phase transition: ferromagnetic or antiferromagnetic.

The change in free energy associated with the third component can be described as the difference between the three-component state ($f(c_{Pt}, c_{Pd},\eta)$) and the binary state ($f(c_{Pt}, \eta)$). In this analysis, we consider the disordered state with the long-range order parameter ($\eta$) set to zero.

\begin{gather}
    \Delta f(c_{Pd}) = f(c_{Pt}, c_{Pd}, 0) - f(c_{Pt}, 0) = 
    \nonumber \\
    \frac{1}{2}  \bigg [ \bigg (
       c^2_{Pd} \tilde{w}_{(Fe-Pt-Pd)}(\textbf{0}) - c^2_{Pt} \tilde{w}_{(Fe-Pt)}(\textbf{0})
       + 2 c_{Fe} c_{Pd}\tilde{J}_{FePd}(\textbf{0}) s_{Fe} \sigma_{Fe} s_{Pd} \sigma_{Pd} 
       \bigg )
        \\
              - k_{B}Tc_{Pd}
                \bigg [lnsh \bigg(
                        \bigg (
                            1 + 
                            \frac{1}{2s_{Pd}}
                        \bigg )
                        y_{Pd}(\sigma_{Pd})
                    \bigg ) \nonumber
                    -
                    lnsh \bigg (\frac{1}{2s_{Pd}}y_{Pd}(\sigma_{Pd}) \bigg)
                    -
                    \sigma_{Pd}y_{Pd}(\sigma_{Pd} ) \bigg]
        \label{free_en_dif_2}
\end{gather}

The increase in Curie temperature due to atomic (positional) ordering can be analytically expressed as the difference in free energy between the ordered ($f(c_{Pt}, c_{Pd}, \eta)$) and disordered ($f(c_{Pt}, c_{Pd}, 0)$) states for Fe-Pt-Pd alloy:
\begin{gather}
    \Delta f(\eta) = f(c_{Pt}, c_{Pd}, \eta) - f(c_{Pt}, c_{Pt}, 0) =
    \nonumber \\
    \frac{3}{16}\eta^2 \bigg (
            \tilde{w}_{prm}(\textbf{k}_x) +
            \tilde{J}_{PtPt}(\textbf{k}_x)s^2_{Pt}\sigma^2_{Pt} 
            \nonumber
            + \tilde{J}_{FeFe}(\textbf{k}_x)
            s^2_{Fe}\sigma^2_{Fe}
            - 2\tilde{J}_{FePt}(\textbf{k}_x)s_{Fe}s_{Pt}\sigma_{Fe}\sigma_{Pt}  
            \\
            + \tilde{J}_{PdPd}(\textbf{k}_x)s^2_{Pd}\sigma^2_{Pd}
            - 2\tilde{J}_{FePd}(\textbf{k}_x)s_{Fe}s_{Pd}\sigma_{Fe}\sigma_{Pd}
            - 2\tilde{J}_{PtPd}(\textbf{k}_x)s_{Pt}s_{Pd}\sigma_{Pt}\sigma_{Pd} \bigg )
            \nonumber \\
            + \frac{k_{B}T}{4}
            \bigg [
                3 \bigg (c_{Pt} - \frac{\eta}{4} \bigg )ln \bigg (c_{Pt} - \frac{\eta}{4} \bigg )
                +
                3 \bigg (c_{Fe} + \frac{\eta}{4} \bigg)ln\bigg (c_{Fe} + \frac{\eta}{4} \bigg) \\
                +
                \bigg (c_{Fe} - \frac{3\eta}{4} \bigg )ln \bigg ( c_{Fe} - \frac{3 \eta}{4} \bigg) \nonumber
                +
                \bigg (c_{Pt} + \frac{3\eta}{4} \bigg)
                ln \bigg (c_{Pt} + \frac{3\eta}{4} \bigg ) - 4c_{Pt}lnc_{Pt} -4c_{Fe}lnc_{Fe} \bigg ) \bigg] 
\end{gather}

Due to the weak magnetic interactions between Pd-Pd and Pt-Pd atoms, we have neglected the terms   
$\tilde{J}_{PdPd}$, $\tilde{J}_{PtPd}$.

The term $\bigg (c^2_{Pd}\tilde{w}_{(Fe-Pt-Pd)}(\textbf{k}_x) - c^2_{Pt}\tilde{w}_{(Fe-Pt)}(\textbf{k}_x) \bigg )$  represents the positional disorder introduced by the addition of the third component (Pd) to the binary Fe-Pt alloy, as described by Madelung \cite{Madelung}.

The term $\frac{3}{16}\eta^2 \bigg (
            \tilde{w}_{(Fe-Pt-Pd)}(\textbf{k}_x) - \tilde{w}_{(Fe-Pt)}(\textbf{k}_x) \bigg)$ 
indicates the atomic order in the Fe-Pt-Pd alloy.

The minimum and maximum of the functions representing the changes in thermodynamic state — due to doping with Pd, ($\Delta f(c_{\text{Pd}})$), and atomic ordering, ($\Delta f(\eta)$) — can be investigated using the following conditions:
\begin{gather}
    \frac{\partial \big(f(c_{Pt}, c_{Pd}, \eta) - f(c_{Pt}, c_{Pd}, 0) \big )}{\partial d\eta} = 0
    \\
    \frac{\partial \big(f(c_{Pt}, c_{Pd}, 0) - f(c_{Pt}, 0) \big )}{\partial dc_{Pd}} = 0 
    \label{first_der}
\end{gather}

\begin{gather}
    \frac{\partial^2 \big(f(c_{Pt}, c_{Pd}, \eta) - f(c_{Pt}, c_{Pd}, 0) \big )}{\partial d\eta^2} > or < 0
    \\
     \frac{\partial^2 \big(f(c_{Pt}, c_{Pd}, 0) - f(c_{Pt}, 0) \big )}{\partial dc_{Pd}^2} > or < 0  
    \label{second_der}
\end{gather}
Since the temperature of magnetic transition demonstrates non-linear behavior (Fig. \ref{fig:1}, Fig. \ref{fig:3}) versus Pd concentration,
these conditions can define a critical palladium concentration where the configuration free energy shows an extremum. It's defined by 'mixing' energy $\tilde{w_{(Fe-Pt-Pd)}}$, exchange coupling constant between Fe and Pd, spin numbers, and the relative spontaneous magnetization.

Since the temperature of the magnetic transition exhibits non-linear behavior with respect to Pd concentration (see Fig. \ref{fig:1} and Fig. \ref{fig:3}), conditions \ref{first_der} -\ref{second_der} can be used to identify a critical palladium concentration at which the configurational free energy reaches an extremum. This critical point is influenced by the 'mixing' energy $\tilde{w}_{(\text{Fe-Pt-Pd})}(\textbf{0})$, the exchange coupling constant between Fe and Pd, the spin numbers, and the relative spontaneous magnetization.

$\tilde{w}_{(Fe-Pt-Pd)}(\textbf{0})$ in condition \ref{second_der} will be obviously less than zero in condition, indicating a maximum of Curie/Néel temperature seen in Fig. \ref{fig:1}, Fig. \ref{fig:3}.
The result for the long-range order parameter is trivial - the maximum effect on temperature of magnetic order is reached at the maximum possible atomic order.
        
\section{Summary}

We applied machine learning techniques to predict the temperature of magnetic transformation in Fe-based, Pt-based ternary alloys.

The Azure Machine Learning framework was employed to perform statistical analysis and develop a predictive model for the Curie temperature of these alloy systems.
Feature selection was carried out using a data-driven approach, with the intrinsic characteristics of Fe–Pt alloys serving as the basis. Model evaluation using Monte Carlo cross-validation revealed that the Voting Ensemble algorithm provided the highest predictive performance. 

In our previous work \cite{Ponomarova_ML}, we developed a feature vector for predicting the Curie temperature of binary alloys, defined as $f(c_{Fe},c_{Pt},Z_{Fe}, Z_{Pt},p,\eta, T{c})$. 
In the present study, this methodology is extended to ternary alloys by expanding the feature set to: $f(c_{Fe}, c_{Pt},c_{Pd},r_{Fe},r_{Pt},r_{Pd},s_{Fe},s_{Pt},s_{Pd},Z_{Fe},Z_{Pt},Z_{Pd},\eta,Tc/T_N)$.

It was found that alloying with up to approximately 1 at.\% Pd affects the Curie temperature in Fe-rich alloys. Further increase in Pd concentration produces no substantial change, as the Curie temperature approaches its asymptotic value. Atomic ordering significantly increases the Curie temperature in $Fe_{3}Pt-Pd$ alloys.
Fe-rich Fe–Pt–Pd ternary alloys exhibit ferromagnetic behavior closely resembling that of the binary $Fe_{3}Pt$ compound.

The Pd-rich alloys exhibit a similar tendency in the change in magnetic transition temperature upon replacing Pt with Pd up to 2 at.\%.  The main difference is the presence of a ferromagnetic phase transition in the disordered state and an antiferromagnetic order in the ordered state.

Experimental measurements of magnetic transition temperatures confirmed the accuracy of the model predictions.

Through a thermodynamic analysis, we examined important phenomenological parameters, including the 'mixing energy' of the alloy and the exchange coupling constants. We also calculated the critical values for the long-range order parameter and the concentration of palladium.

\section{Model applications}
The feature set developed in this work demonstrates strong potential for broader applicability across various materials design tasks and multicomponent alloy systems.

Initially, we employed the feature set $f(c_{Fe},c_{Pt},Z_{Fe}, Z_{Pt},p,\eta, T{c})$ created by us for binary systems in \cite{Ponomarova_ML}  and then extended it for three-component alloys by transforming into $f(c_{Fe}, c_{Pt},c_{Pd},r_{Fe},r_{Pt}, r_{Pd},s_{Fe},s_{Pt},s_{Pd},Z_{Fe},Z_{Pt},Z_{Pd},\eta, T_c/T_N)$ where additionally the atomic order parameter $\eta$ and the concentration of the third element (Pd) are used as design variables for tuning the Curie/Néel temperature ($T_c$/$T_N$). 

The feature set can also be applied:

\begin{itemize}
    \item \textit{for other three component alloys (not Fe-Pt-Pd) which undergo atomic ordering.} 
    
    For example, the model can be applied to Fe–Pd-, Fe–Ni-, Co–(Pt,Ni)-, Ni–(Pt,Al) - based, and other alloys that exhibit atomic ordering.
    As an illustrative case, consider the Ni–Al–Pt alloy.
    The corresponding feature set will be transformed from:
    \newline
    $f(c_{Fe}, c_{Pt},c_{Pd},r_{Fe},r_{Pt}, r_{Pd},s_{Fe},s_{Pt},s_{Pd},Z_{Fe},Z_{Pt},Z_{Pd},\eta,T_c/T_N)$
    \newline
    into:
    \newline
    $f(c_{Ni}, c_{Al},c_{Pt},r_{Ni},r_{Al}, r_{Pt},s_{Ni},s_{Al},s_{Pt},Z_{Ni},Z_{Al},Z_{Pt},\eta,T_c/T_N)$
    
    \item \textit{for three- four four-component alloys under the external magnetic field ($H$)}.
    
    Let's take Fe-Pt-Pd alloy as an illustrative example. The feature set example will be transformed from:
    \newline
    $f(c_{Fe}, c_{Pt},c_{Pd},r_{Fe},r_{Pt}, r_{Pd},s_{Fe},s_{Pt},s_{Pd},Z_{Fe},Z_{Pt},Z_{Pd},\eta,T_c/T_N)$
    \newline
    into:
    \newline
    $f(c_{Fe}, c_{Pt},c_{Pd},r_{Fe},r_{Pt}, r_{Pd},s_{Fe},s_{Pt},s_{Pd},Z_{Fe},Z_{Pt},Z_{Pd},\eta,H,T_c/T_N)$
    \item \textit{for three and four-component alloys under external pressure ($p$)}. 
    \newline
    Let's take the same Fe-Pt-Pt alloy under external pressure. 
    The feature set will be transformed from:
    \newline
    $f(c_{Fe}, c_{Pt},c_{Pd},r_{Fe},r_{Pt}, r_{Pd},s_{Fe},s_{Pt},s_{Pd},Z_{Fe},Z_{Pt},Z_{Pd},\eta,T_c/T_N)$
    \newline
    into:
    \newline
    $f(c_{Fe}, c_{Pt},c_{Pd},r_{Fe},r_{Pt}, r_{Pd},s_{Fe},s_{Pt},s_{Pd},Z_{Fe},Z_{Pt},Z_{Pd},\eta,p,T_c/T_N)$
    
    \item \textit{for high-entropy alloys}. 
    \newline
    Usually, high-entropy alloys are multicomponent with X components, then the feature set will be generalized to X concentrations, X component radii, X spin numbers, and X atomic numbers.
    In the case of a four-component HE alloy (e.g., Fe–Ni–Co–Mn), the feature set will be transformed from: 
    \newline
    $f(c_{Fe}, c_{Pt},c_{Pd},r_{Fe},r_{Pt}, r_{Pd},s_{Fe},s_{Pt},s_{Pd},Z_{Fe},Z_{Pt},Z_{Pd},\eta,T_c/T_N)$
    \newline
    into
    \newline
$f(c_{Fe},c_{Ni},c_{Co},c_{Mn},r_{Fe},r_{Ni},r_{Co},r_{Mn},Z_{Fe},Z_{Ni}, Z_{Co}, Z_{Mn}, s_{Fe}, s_{Ni}, s_{Co}, s_{Mn},T_c/T_N)$
    \item \textit{for magnetic nanoparticles}. 
    As is known, the size effect is a significant design factor for magnetic nanoparticles. Adding the size of the nanoparticle ($R$) and considering the shape factor ($\alpha$) in the feature set application may be extended for nanoobjects. 
    
    The feature set example for a three-component alloy (let's take Fe-Pt-Pd to illustrate) will be transformed from:
    \newline
    $f(c_{Fe}, c_{Pt},c_{Pd},r_{Fe},r_{Pt}, r_{Pd},s_{Fe},s_{Pt},s_{Pd},Z_{Fe},Z_{Pt},Z_{Pd},\eta,T_c/T_N)$
    \newline
    into:
    \newline
    $f(c_{Fe}, c_{Pt},c_{Pd},r_{Fe},r_{Pt}, r_{Pd},s_{Fe},s_{Pt},s_{Pd},Z_{Fe},Z_{Pt},Z_{Pd},\eta,R,\alpha,T_c/T_N)$
\end{itemize}

\section{Conflict of interest}
The authors declare no conflict of interest.

\section{Acknowledgment}
This paper would not have been possible without the exceptional support of Alla Lysak.

\section{Data Availability Statement}
Data is available per reasonable request.

\section{Funding details}
No special funding was provided for this work.

\section{Disclosure statement}
The authors report there are no competing interests to declare.

\medskip
\bibliography{bibfile}

\end{document}